\documentclass[submission,copyright,creativecommons]{eptcs}
\providecommand{\event}{Wivace 2013} 
\usepackage{breakurl}             

\usepackage{amsfonts}
\usepackage{amssymb}
\usepackage{graphicx}
\usepackage{amsmath}
\usepackage{rotating}
\usepackage{tabularx}
\usepackage{setspace} 
\usepackage{color}

\title{Self Organizing Maps to efficiently cluster and functionally interpret protein conformational ensembles}

\author{Domenico Fraccalvieri* \textsuperscript{1} \qquad Laura Bonati \textsuperscript{1} \qquad Fabio Stella \textsuperscript{2}
\institute{\textsuperscript{1}Department of Earth and Environmental Sciences, University of Milano Bicocca, Milano, IT\\
	\textsuperscript{2}Department of Informatics, Systems and Communication, University of Milano Bicocca, Milano, IT}
\email{*Corresponding Author = domenico.fraccalvieri@unimib.it}
}

\def\titlerunning{SOMs to efficiently cluster and functionally interpret protein conformational ensembles}
\def\authorrunning{D. Fraccalvieri, L. Bonati \& F. Stella}
\begin{document}
\maketitle


\section{Introduction}
The wide range of protein biological functions, such as enzymatic activity, ligand- and protein-protein interactions and allosteric regulation, is strictly related to their flexibility and dynamics\cite{Boehr:2009:Nat-Chem-Biol:19841628}. To model the influence of protein motions across this broad spectrum of events Molecular Dynamics (MD) simulations are now routinely used. The identification of the most functionally relevant conformations is generally done by grouping the conformations according to a criterion of geometrical similarity and popular choices include hierarchical clustering, single, complete and average linkage and k-means \cite{ISI:000251024200037}. These geometrical approaches rely on the assumption that the identified conformational states also correspond to the energetic states \cite{ISI:000251024200037}. Good candidates to improve this match are Artificial Neural Networks (ANNs) which are capable to discover the relationships between the measured variables only from the available dataset \cite{10281_43815}. Among the class of ANNs, successfully applied to artificial life systems \cite{cangelosi2003artificial}, Self Organizing Maps (SOMs) \cite{kohonen-2012} represent a particularly powerful data driven model that has been widely applied for exploration and clustering of high-dimensional datasets \cite{10281_43815,kohonen-2012}. 
We recently developed an approach which combines SOMs and hierarchical clustering to efficiently compare conformational ensembles obtained from multiple MD simulations of proteins \cite{journals/bmcbi/FraccalvieriPSB11}. To reliably apply the SOM analysis to these specific input data we identified and optimized a small number of SOM parameters. In particular, we confirmed that the map size is a crucial parameter and that a well-selected number of neurons is crucial both to reduce the computational cost of the analysis and to provide an intermediate topological representation of the input conformational space \cite{journals/bmcbi/FraccalvieriPSB11}. As a result the original MD trajectories can be represented by the few conformations that best represent the clusters obtained.
Here we make a step further; the proposed approach consists of processing all the atom positions of each conformation won by a given SOM neuron using a similarity network. Different similarity measures between atoms behavior are used to compile a similarity matrix which is inputted to a network model \cite{Newman2010}. Network algorithms are then used to automatically discover and interpret the behavior of the original protein conformational ensembles exploiting the atomic coordinates enclosed in the SOM neuron.

\section{Methods}
\textbf{Self Organizing Map clustering.}  
The details of the proposed SOM approach can be summarized as follow. First, an ensemble of conformations is extracted from one (or more) MD trajectory with an optimized sampling rate of 1/100 ps. Each conformation of the trajectories is represented using only the C$\alpha$ Cartesian coordinates. The ensemble is encoded in a matrix and is used for the learning stage, performed with the set of optimal SOM parameters previously identified by experimental design \cite{journals/bmcbi/FraccalvieriPSB11}. The most relevant parameters are: \emph{Map size=100, Radius=3, Training length=5000 and Neighbor function=gaussian}. The neurons of this “trained SOM”, i.e. the proto-clusters of the original conformational ensemble, are grouped using the complete linkage algorithm and the number of clusters is automatically selected by means of the Mojena's stopping rule \cite{journals/cj/Mojena77}. To provide a graphical representation of the neurons and their connections, each neuron is described by a hexagon and the neurons belonging to each cluster are homogeneously colored (see Figure 1, central panel). 
Finally, to provide the most synthetic structural representation of the original ensemble, only the input conformation closest to the centroid of each cluster is represented (see Figure 1, left panel). The centroid is defined as the mean vector of all the prototype vectors of the neurons in that cluster.\\
\\
\textbf{Network Analysis.}  
The set of all the C$\alpha$'s atomic coordinates of the conformations won by each SOM neuron, are used to compile a network model. The following similarity measures were analyzed and empirically compared; x-y-z average absolute correlation, cosine, Pearson, Spearman correlation and biweight mid-correlation. Networks are constructed from similarity matrices by means of threshold as described in \cite{horvath2011weighted}. The SOM, which summarizes the atomic positions of each conformation associated with a neuron, allows building a network model which is then exploited to automatically discover and interpret the behavior of the inputted protein conformational ensembles. Graph partitioning algorithms \cite{horvath2011weighted} to cluster atoms with similar behavior and community detection algorithms \cite{Newman2010,horvath2011weighted} were applied to the network model (see Figure 1, right panel).

\begin{figure}[b!]
\begin{center}
	\includegraphics[height=16cm, width=16cm, keepaspectratio]{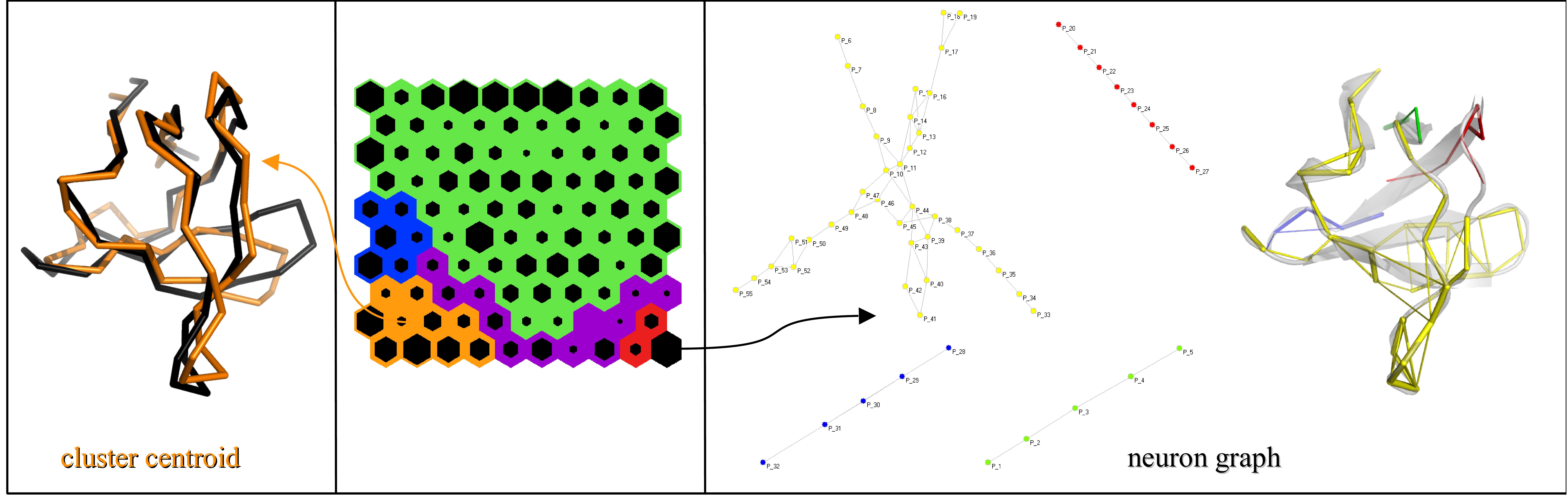}
	\caption{Central panel: the “clustered SOM”. The background colors of the neurons (hexagons) correspond to the cluster attribution. The inner hexagon is proportional to the number of conformations won by the neuron. Left panel: ribbon representation of one centroid cluster (in orange) superimposed to a reference structure (in black). Right panel: graphical analysis of one neuron. On the left, each node represents an atom of the neuron conformations and the edges link pairs of correlated atoms. The groups of nodes (atoms) with similar behavior are homogeneously colored. On the right, the nodes and the edges are represented on a reference 3D structure (in gray) using the same color code of the graph. The width of the edge is proportional to the correlation among the atoms (nodes).}
	\label{SOM_results}
\end{center}
\end{figure}

\section{Results and Discussion}
The SOM approach developed by our group has been successfully applied showing its potential in the comparison of MD simulations of different proteins with dimensions ranging from 60 to 500 amino acids\cite{journals/bmcbi/FraccalvieriPSB11,C2MB25192B}. These applications highlighted key features, related to the topological nature of the SOM. First, a single map was able to detect both low fluctuations, i.e. little conformational changes with respect to a reference structure, and large concerted motions, i.e. simultaneous conformational changes of different region of the protein.\cite{journals/bmcbi/FraccalvieriPSB11,C2MB25192B}. Second, the intrinsic time dependence of the data, not given as input, was correctly recovered; in fact consecutive conformational transitions were correctly assigned to adjacent neurons, thus to adjacent clusters, on the map  \cite{journals/bmcbi/FraccalvieriPSB11}.
Moreover, the values of geometrical descriptors based on a subset of C$\alpha$ atoms included in the SOM training \cite{journals/bmcbi/FraccalvieriPSB11,C2MB25192B}, as well as excluded features  \cite{C2MB25192B}, were consistently clustered, indicating the potential of the map to classify unknown features  \cite{C2MB25192B}. Finally the ability of a SOM trained with a set of protein to classify conformations of similar protein was verified and exploited \cite{C2MB25192B}.
The extension of our protocol here presented is aimed at a deeper inspection of the dynamic information contained in each neuron in the output map. In fact, not only the subset of conformations won by each neuron, but also the specific motions caught by that neuron are analyzed. To this end, the matrices composed by the C$\alpha$ Cartesian coordinates of the original data in each neuron are analyzed using similarity matrices and represented by graphs (see Figure 1, right panel). This analysis has a double outcome. First it provides an interesting and informative picture of the joint behavior of the atoms belonging to the conformations in each neuron, highlighting both local motions, i.e. motions involving structurally neighbor atoms, and long-range communications mediated by atoms able to transmit conformational signals to remote sites of the protein. Second this information can be used to cluster the data according to this graph. Cluster analysis based on  similarity of the C$\alpha$ dynamic behavior enclosed in each SOM neuron should increase the ability of our approach to compare the dynamics of different proteins.
In conclusion, we already showed that the use of this SOM approach to cluster protein conformational ensembles increases the cluster quality compared to other methods routinely used \cite{journals/bmcbi/FraccalvieriPSB11}. 
Moreover both the approaches proposed for the map interpretation allow an efficient comparison of the dynamics of different proteins. In the former approach, the characteristics of the conformations belonging to each cluster are summarized and described by the cluster centroid, in the latter, the characteristic pathways of motion described by each neuron are highlighted by the graphs (Figure 1). 
Applications of this approach should range from protein engineering (to design specific mutation able to reduce, or increase, the function of a protein) to computer-based drug design (to select specific conformations of a protein suitable to interact with specific ligands).


\nocite{*}
\bibliographystyle{eptcs}
\bibliography{FraBonSte}

\begin{thebibliography}{10}
\providecommand{\bibitemdeclare}[2]{}
\providecommand{\surnamestart}{}
\providecommand{\surnameend}{}
\providecommand{\urlprefix}{Available at }
\providecommand{\url}[1]{\texttt{#1}}
\providecommand{\href}[2]{\texttt{#2}}
\providecommand{\urlalt}[2]{\href{#1}{#2}}
\providecommand{\doi}[1]{doi:\urlalt{http://dx.doi.org/#1}{#1}}
\providecommand{\bibinfo}[2]{#2}

\bibitemdeclare{article}{Boehr:2009:Nat-Chem-Biol:19841628}
\bibitem{Boehr:2009:Nat-Chem-Biol:19841628}
\bibinfo{author}{D.D. \surnamestart Boehr\surnameend},
  \bibinfo{author}{R.~\surnamestart Nussinov\surnameend} \&
  \bibinfo{author}{P.E. \surnamestart Wright\surnameend}
  (\bibinfo{year}{2009}): \emph{\bibinfo{title}{The role of dynamic
  conformational ensembles in biomolecular recognition}}.
\newblock {\sl \bibinfo{journal}{Nat. Chem. Biol.}}
  \bibinfo{volume}{5}(\bibinfo{number}{11}), pp. \bibinfo{pages}{789--796},
  \doi{10.1038/nchembio.232}.

\bibitemdeclare{inproceedings}{cangelosi2003artificial}
\bibitem{cangelosi2003artificial}
\bibinfo{author}{A.~\surnamestart Cangelosi\surnameend}, \bibinfo{author}{Nolfi
  \surnamestart S.\surnameend} \& \bibinfo{author}{Parisi \surnamestart
  D.\surnameend} (\bibinfo{year}{2003}): \emph{\bibinfo{title}{Artificial Life
  Models of Neural Development}}.
\newblock In: {\sl \bibinfo{booktitle}{On Growth, form and Computers}},
  \bibinfo{publisher}{Elsevier Academic Press}, pp. \bibinfo{pages}{339--52}.
\doi{10.1016/B978-012428765-5/50051-7}

\bibitemdeclare{article}{journals/bmcbi/FraccalvieriPSB11}
\bibitem{journals/bmcbi/FraccalvieriPSB11}
\bibinfo{author}{D.~\surnamestart Fraccalvieri\surnameend},
  \bibinfo{author}{A.~\surnamestart Pandini\surnameend},
  \bibinfo{author}{F.~\surnamestart Stella\surnameend} \&
  \bibinfo{author}{L.~\surnamestart Bonati\surnameend} (\bibinfo{year}{2011}):
  \emph{\bibinfo{title}{Conformational and functional analysis of molecular
  dynamics trajectories by Self-Organising Maps.}}
\newblock {\sl \bibinfo{journal}{BMC Bioinformatics}} \bibinfo{volume}{12}, p.
  \bibinfo{pages}{158}, \doi{10.1186/1471-2105-12-158}.

\bibitemdeclare{article}{C2MB25192B}
\bibitem{C2MB25192B}
\bibinfo{author}{D.~\surnamestart Fraccalvieri\surnameend},
  \bibinfo{author}{M.~\surnamestart Tiberti\surnameend},
  \bibinfo{author}{A.~\surnamestart Pandini\surnameend},
  \bibinfo{author}{L.~\surnamestart Bonati\surnameend} \&
  \bibinfo{author}{E.~\surnamestart Papaleo\surnameend} (\bibinfo{year}{2012}):
  \emph{\bibinfo{title}{Functional annotation of the mesophilic-like character
  of mutants in a cold-adapted enzyme by self-organising map analysis of their
  molecular dynamics}}.
\newblock {\sl \bibinfo{journal}{Mol. Biosyst.}} \bibinfo{volume}{8}, pp.
  \bibinfo{pages}{2680--91}, \doi{10.1039/C2MB25192B}.

\bibitemdeclare{book}{horvath2011weighted}
\bibitem{horvath2011weighted}
\bibinfo{author}{S.~\surnamestart Horvath\surnameend} (\bibinfo{year}{2011}):
  \emph{\bibinfo{title}{Weighted Network Analysis: Applications in Genomics and
  Systems Biology}}.
\newblock \bibinfo{publisher}{Springer}.
\doi{10.1007/978-1-4419-8819-5}

\bibitemdeclare{article}{kohonen-2012}
\bibitem{kohonen-2012}
\bibinfo{author}{T.K. \surnamestart Kohonen\surnameend} (\bibinfo{year}{2013}):
  \emph{\bibinfo{title}{{Essentials of the Self-Organizing Map}}}.
\newblock {\sl \bibinfo{journal}{Neural Networks}} \bibinfo{volume}{37}, pp.
  \bibinfo{pages}{52--65}, \doi{10.1016/j.neunet.2012.09.018}.

\bibitemdeclare{article}{journals/cj/Mojena77}
\bibitem{journals/cj/Mojena77}
\bibinfo{author}{R.~\surnamestart Mojena\surnameend} (\bibinfo{year}{1977}):
  \emph{\bibinfo{title}{Hierarchical Grouping Methods and Stopping Rules: An
  Evaluation.}}
\newblock {\sl \bibinfo{journal}{Comput. J.}}
  \bibinfo{volume}{20}(\bibinfo{number}{4}), pp. \bibinfo{pages}{359--63},
  \doi{10.1093/comjnl/20.4.359}.

\bibitemdeclare{book}{Newman2010}
\bibitem{Newman2010}
\bibinfo{author}{M.E.J. \surnamestart Newman\surnameend}
  (\bibinfo{year}{2010}): \emph{\bibinfo{title}{Networks: An Introduction}}.
\newblock \bibinfo{publisher}{Oxford University Press}, \bibinfo{address}{New
  York}.

\bibitemdeclare{article}{10281_43815}
\bibitem{10281_43815}
\bibinfo{author}{A.~\surnamestart Pandini\surnameend},
  \bibinfo{author}{D.~\surnamestart Fraccalvieri\surnameend} \&
  \bibinfo{author}{L.~\surnamestart Bonati\surnameend} (\bibinfo{year}{2013}):
  \emph{\bibinfo{title}{Artificial Neural Networks for Efficient Clustering of
  Conformational Ensembles and their Potential for Medicinal Chemistry}}.
\newblock {\sl \bibinfo{journal}{Curr. Top. Med. Chem.}}
  \bibinfo{volume}{13}(\bibinfo{number}{5}), pp. \bibinfo{pages}{642--51},
  \doi{10.2174/1568026611313050007}.

\bibitemdeclare{article}{ISI:000251024200037}
\bibitem{ISI:000251024200037}
\bibinfo{author}{J.~\surnamestart Shao\surnameend}, \bibinfo{author}{S.W.
  \surnamestart Tanner\surnameend}, \bibinfo{author}{N.~\surnamestart
  Thompson\surnameend} \& \bibinfo{author}{T.E. \surnamestart
  Cheatham\surnameend} (\bibinfo{year}{{2007}}):
  \emph{\bibinfo{title}{{Clustering molecular dynamics trajectories: 1.
  Characterizing the performance of different clustering algorithms}}}.
\newblock {\sl \bibinfo{journal}{{J. Chem. Theory Comput.}}}
  \bibinfo{volume}{{3}}(\bibinfo{number}{{6}}), pp.
  \bibinfo{pages}{{2312--34}}, \doi{{10.1021/ct700119m}}.

\end{thebibliography}
\end{document}